\documentclass[useAMS,usenatbib]{mn2e}
\usepackage{natbib}
\usepackage{amsmath,amsfonts}
\usepackage{epsfig}
\usepackage{mathptmx}
\usepackage{textcomp}
\usepackage{amssymb} 
\usepackage{natbib}
\usepackage{color}

\def\reff@jnl#1{{\rm#1\/}}

\def\aj{\reff@jnl{AJ}}                  
\def\araa{\reff@jnl{ARA\&A}}            
\def\apj{\reff@jnl{ApJ}}                        
\def\apjl{\reff@jnl{ApJ}}               
\def\apjs{\reff@jnl{ApJS}}              
\def\ao{\reff@jnl{Appl.Optics}}         
\def\apss{\reff@jnl{Ap\&SS}}            
\def\aap{\reff@jnl{A\&A}}               
\def\aapr{\reff@jnl{A\&A~Rev.}}         
\def\aaps{\reff@jnl{A\&AS}}             
\def\azh{\reff@jnl{AZh}}                        
\def\baas{\reff@jnl{BAAS}}              
\def\jrasc{\reff@jnl{JRASC}}            
\def\memras{\reff@jnl{MmRAS}}           
\def\mnras{\reff@jnl{MNRAS}}            
\def\pra{\reff@jnl{Phys. Rev. A}}         
\def\prb{\reff@jnl{Phys. Rev. B}}         
\def\prc{\reff@jnl{Phys. Rev. C}}         
\def\prd{\reff@jnl{Phys. Rev. D}}         
\def\prl{\reff@jnl{Phys. Rev. Lett}}      
\def\pasp{\reff@jnl{PASP}}              
\def\pasj{\reff@jnl{PASJ}}              
\def\qjras{\reff@jnl{QJRAS}}            
\def\rmxaa{\reff@jnl{RMxAA}}		
\def\skytel{\reff@jnl{S\&T}}            
\def\solphys{\reff@jnl{Solar~Phys.}}    
\def\sovast{\reff@jnl{Soviet~Ast.}}     
\def\ssr{\reff@jnl{Space~Sci.Rev.}}     
\def\zap{\reff@jnl{ZAp}}                        
\def\nat{\reff@jnl{Nature}}             

\def\p#1by#2{{\partial{#1} \over \partial{#2}}}
\def\pp#1by#2#3{{\partial^2{#1} \over \partial{#2}\partial{#3}}}
\def\d#1by#2{{{\rm d}{#1} \over {\rm d}{#2}}}
\def\dd#1by#2#3{{{\rm d}^2{#1} \over {\rm d}{#2}{\rm d}{#3}}}

\title[DG~Tau with e-MERLIN]{Sub-arcsecond high sensitivity measurements of the DG~Tau jet with e-MERLIN}
\label{firstpage}

\author[Ainsworth et~al.]{
 Rachael E. Ainsworth$^{1}$\thanks{email: rainsworth@cp.dias.ie}, 
 Tom P. Ray$^{1}$, 
 Anna M. M. Scaife$^{2},$
 Jane S. Greaves$^{3}$ \&
 Rob J. Beswick$^{4}$.
 \vspace{0.03in}\\
$^1$ Dublin Institute for Advanced Studies, 31 Fitzwilliam Place, Dublin 2, Ireland\\
$^2$ School of Physics \& Astronomy, University of Southampton, Highfield, Southampton, SO17 1BJ\\
$^3$ SUPA, School of Physics and Astronomy, University of St. Andrews, North Haugh, St. Andrews, Fife KY16 9SS\\
$^4$ Jodrell Bank Center for Astrophysics, The University of Manchester, Oxford Rd, Manchester M13 9PL, UK
}

\date{Accepted 2013 August 09; received 2013 August 08; in original form 2013 May 13}

\pagerange{\pageref{firstpage}--\pageref{lastpage}}

\pubyear{2012}

\begin{document}
\maketitle

\begin{abstract}
We present very high spatial resolution deep radio continuum observations at 5\,GHz (6\,cm) made with e-MERLIN of the young stars DG~Tau~A and B. Assuming it is launched very close ($\simeq1$\,au) from the star, our results suggest that the DG~Tau~A outflow initially starts as a poorly focused wind and undergoes significant collimation further along the jet ($\simeq50$\,au). We derive jet parameters for DG~Tau~A and find an initial jet opening angle of $86\degr$ within 2\,au of the source, a mass-loss rate of $1.5\times10^{-8}$\,M$_{\sun}$\,yr$^{-1}$ for the ionised component of the jet, and the total ejection/accretion ratio to range from $0.06-0.3$. These results are in line with predictions from MHD jet-launching theories.

\end{abstract}


\section{Introduction}
\label{intro}

Young stellar objects (YSOs) drive powerful outflows due to accretion from a surrounding envelope and circumstellar disc. The Class~II stage of low-mass protostellar evolution, also known as the Classical T~Tauri Star (CTTS) stage, occurs when most of the original core has been accreted, resulting in lower accretion rates through the surrounding accretion disc onto the YSO and the system becomes optically visible \citep[e.g.][]{2007prpl.conf..277P}. The physical mechanism by which these outflows are launched and collimated into jets still remains a mystery, however it is widely accepted that magnetic fields play an important role \citep[e.g.][]{2007IAUS..243..203C}. The competing theories for a magnetohydrodynamic (MHD) jet launching mechanism are the X-wind model \citep{2000prpl.conf..789S}, where the jet originates from the interface between the star's magnetosphere and disc, and the disc-wind model \citep{2006A&A...453..785F, 2007prpl.conf..277P}, where the jet is launched from disc radii of 0.1 to a few au. A stellar component may also be present \citep{2012ApJ...745..101M}. CTTSs are a perfect laboratory for studying the jet launching and collimation mechanisms as their proximity (140\,pc to the nearest star forming regions) and lack of obscuring envelope give access to the inner jet regions within 20-200\,au, where the outflow structure might not yet be significantly disturbed by interaction with the ambient medium \citep{2007IAUS..243..183R, 2011A&A...532A..59A}. 

Observations with radio interferometers of YSO outflows have the advantage that their spatial resolution is comparable or better than the Hubble Space Telescope. Moreover, the new class of radio interferometers, such as the extended Multi-Element Radio Linked Interferometer Network (e-MERLIN) in the UK and the Jansky Very Large Array (JVLA) in the US, have vastly improved sensitivity to detect the central jet engine. The emission at these wavelengths for YSOs normally arises from thermal bremsstrahlung radiation \citep[e.g.][]{2012MNRAS.423.1089A}. The emission commonly has a flat or positive power-law spectral index $\alpha$, where the flux density $S_{\nu}\propto\nu^{\alpha}$ at frequency $\nu$, and ranges between $-0.1$ for the optically thin and $+2$ for the optically thick case. However, in a number of YSOs non-thermal emission is also seen from an outflow \citep[e.g.][]{2007IAUS..243..183R, 2010Sci...330.1209C} with spectral indices that suggest gyrosynchrotron or synchrotron radiation. 

DG~Tau~A is a highly active CTTS located at a distance of 140\,pc in the Taurus Molecular Cloud, and was one of the first T~Tauri stars to be associated with an optical jet \citep[HH~158,][]{1983ApJ...274L..83M}. The optical outflow is observed to have an onion-like kinematic structure within 500\,au of the star, with faster and more collimated gas bracketed by wider and slower material, and the flow becomes gradually denser and of higher excitation close to the central axis \citep{2000ApJ...537L..49B}. This behaviour is naturally expected if the wind is launched from a broad range of disc radii \citep{2011A&A...532A..59A}. Signatures of jet rotation have been observed \citep{2002ApJ...576..222B, 2004ApJ...604..758C, 2007ApJ...663..350C}, supporting a magneto-centrifugal jet launch scenario \citep{2007prpl.conf..277P} and the disc has also been shown to rotate in the same direction as the jet \citep{2002A&A...394L..31T}. The HH~158 jet has a position angle (PA) of $223\degr$ \citep[][]{2002ApJ...576..222B} with an inclination angle $i\approx38\degr$ w.r.t. the line of sight \citep{1998AJ....115.1554E}, and has been traced out to a total projected distance of $\approx0.5$\,pc \citep{2007A&A...467.1197M}. Radial velocities in the jet have been found to range up to $\sim350$\,km\,s$^{-1}$ with average velocities of $\sim200$\,km\,s$^{-1}$ \citep{2000A&A...357L..61D}. Extended X-ray emission has also been detected along the outflow \citep{2007A&A...468..515G, 2008A&A...478..797G}. 

Located $\approx55\arcsec$ to the southwest of DG~Tau~A is DG~Tau~B, the driving source of the asymmetrical optical jet HH~159 \citep{1983ApJ...274L..83M} and not thought to be related to DG~Tau~A except by projected proximity \citep{1986ApJ...311L..23J}. DG~Tau~B is a low-luminosity, deeply embedded Class~I source \citep[$L_{\rm bol}=0.7\,L_{\sun}$,][]{2004ApJS..154..391W}. It drives a narrow optical jet \citep{1991A&A...252..740M} and a collimated, redshifted CO outflow \citep{1994ApJ...436L.177M}, both of which are well aligned at a PA of $\approx294\degr$. PAs between $116-122\degr$ have been found for the blueshifted jet \citep{2004A&A...420..975M, 1991A&A...252..740M}. The total projected length of the DG~Tau~B bipolar outflow is $\approx0.5$\,pc \citep{2004A&A...420..975M}. A circumstellar disc has been observed at mm-wavelengths and is perpendicular to the outflow axis \citep{2000ApJ...529..477L}.

In the radio, both DG~Tau~A and B have compact and elongated morphologies in the known direction of their outflows \citep{1982ApJ...253..707C, 1984ApJ...282..699B, 2012A&A...537A.123R} and possess a positive spectral index $\alpha$ typical of free-free emission. Radio spectral energy distributions (SEDs) for these sources are presented in \citet{2012MNRAS.420.3334S} and \citet{Lynch2013} and show evidence for variability. 

In this letter we present the highest resolution radio images of DG~Tau~A and B made to date and use these new data to constrain the jet opening angle and mass-loss rate in DG~Tau~A. In Section~\ref{sec:obs} we present details of these observations and our results in Section~\ref{sec:results}. In Section~\ref{sec:disc} we discuss the implications of our findings, compare them with previous data and mock observations, and derive jet parameters. In Section~\ref{sec:conc} we summarise our conclusions.

\section{Observations}
\label{sec:obs}

The e-MERLIN synthesis telescope is a seven element interferometer with baselines of up to 217\,km and connected by a new optical fibre network to Jodrell Bank Observatory near Manchester, UK. An inhomogeneous array, e-MERLIN is comprised of the 76\,m Lovell telescope, a 32\,m dish at Cambridge, and the following 25\,m antennas: Mark~II, Knockin, Defford, Pickmere and Darnhall. The data in this work were taken during the commissioning phase of e-MERLIN with only the five 25\,m dishes, resulting in a primary beam of $\theta\approx10\arcmin$, a maximum baseline of 133.7\,km (between Pickmere and Defford), and a minimum baseline of 11.2\,km (between Pickmere and Mark II). The final, fully expanded array will have a bandwidth of 2\,GHz, providing more than $10\times$ the continuum sensitivity as the original MERLIN. 

DG~Tau~A was observed at 5\,GHz for $\approx22$\,hrs by e-MERLIN between 1 and 4 August 2011 (average epoch 2011.58), with $4\times128$\,MHz sub-bands of 512 spectral channels each, yielding a total bandwidth of 512\,MHz. Data were reduced and analysed using the National Radio Astronomy Observatory's (NRAO) Astronomical Image Processing System ({\sc AIPS}). Data were initially edited with {\sc spflg} and {\sc ibled}, averaged to 64 channels per intermediate frequency (IF) channel, and concatenated with {\sc dbcon} before further editing was conducted. {\sc fring} was used to derive delay and rate corrections for the calibrator sources, and {\sc calib} was used to derive time-dependent phase, and then amplitude and phase solutions.

Flux calibration was performed using short observations of 3C286 at the beginning and end of each run, and the flux density scale was calculated with {\sc setjy} \citep{2013ApJS..204...19P}. The flux density for each IF was then reduced by 4~percent to account for the resolution of the e-MERLIN shortest spacing (see The MERLIN User Guide\footnote{www.e-merlin.ac.uk}). Calibrator flux densities are listed in Table~\ref{tab:cals}. Bandpass calibration was performed using the bright point sources J\,0555+398 and J\,1407+284, and the phase was calibrated using interleaved observations of J\,0429+2724. Visibilities were re-weighted to take into account the different sensitivity of each dish, and the target source data was {\sc split} for imaging with the calibration applied. Absolute calibration is expected to be accurate to about 10\% for commissioning data. 

\begin{table}
\begin{center}
\caption{e-MERLIN frequency channels and calibrator flux densities measured in Jy.\label{tab:cals}}
\begin{tabular}{lccccc}
\hline\hline
Channel No. & 1 & 2 & 3 & 4 \\
Freq. [GHz] & 4.41 & 4.54 & 4.70 & 4.80 \\
\hline
3C286 & 7.63 & 7.50 & 7.37 & 7.25 \\
J\,0555+398 & 5.46 & 5.47 & 5.36 & 5.55 \\
J\,1407+284 & 2.44 & 2.43 & 2.43 & 2.38 \\
\hline\hline
\end{tabular}
\end{center}
\end{table}

\section{Results}
\label{sec:results}

\begin{figure}
\centerline{\includegraphics[width=0.4\textwidth]{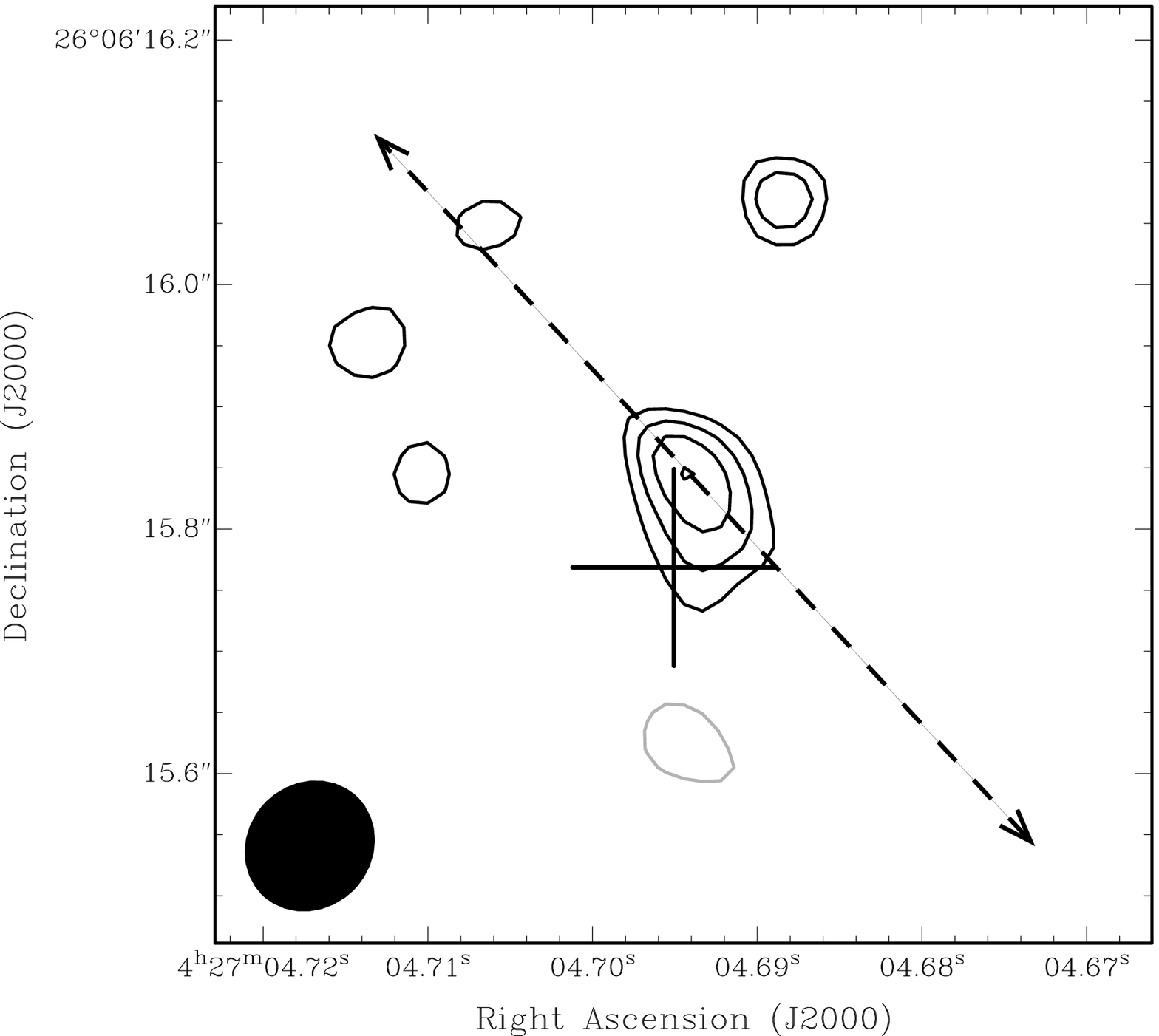}}
\centerline{(a)}
\centerline{\includegraphics[width=0.4\textwidth]{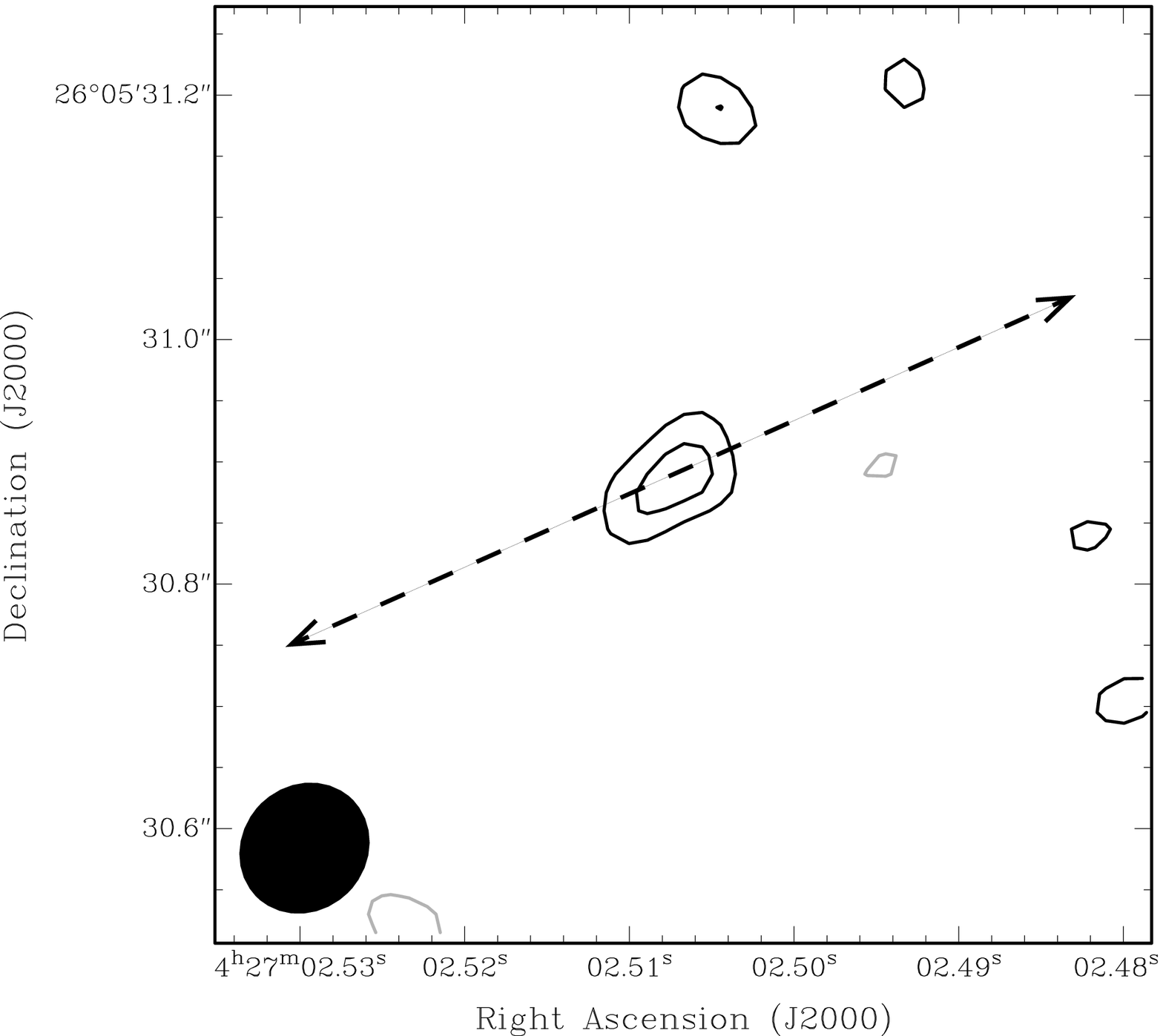}}
\centerline{(b)} 
\caption{e-MERLIN maps at 5\,GHz for (a) DG~Tau~A and (b) DG~Tau~B with contours at -3 (grey), 3, 4, 5, and 6\,$\sigma_{\rm{rms}}$, where $\sigma_{\rm{rms}} = 24\,\mu$Jy\,beam$^{-1}$ for DG~Tau~A and $25\,\mu$Jy\,beam$^{-1}$ for DG~Tau~B. The known outflow direction in both cases is shown as a dashed line, and the optical position for DG~Tau~A is shown as a cross (see Section~\ref{sec:results}). \label{fig:emerlinmap}}
\end{figure}

Deconvolution and imaging were performed with {\sc imagr}. Naturally weighted visibilities were used to ensure optimal signal-to-noise ratio (SNR) levels. Primary beam correction was applied with {\sc pbcor}. The combined channel map centred at 4.67\,GHz (hereafter referred to as 5\,GHz) for DG~Tau~A is shown in Fig.~\ref{fig:emerlinmap}a, and we also detect DG~Tau~B at $4\sigma_{\rm rms}$ within our primary beam (see Fig.~\ref{fig:emerlinmap}b). The dimensions of the synthesised beam are $0.11\times0.10$\,arcsec with a PA of $317.1\degr$. The root-mean-square (rms) noise was measured in {\sc aips} using {\sc imean}, and the values are $24\,\mu$Jy\,beam$^{-1}$ for the DG~Tau~A map and $25\,\mu$Jy\,beam$^{-1}$ for the DG~Tau~B map. 

At Epoch 2011.58, we detect DG~Tau~A at J2000 coordinates $\alpha=04^{\rm{h}} 27^{\rm{m}} 04\fs 693, \delta=+26\degr 06\arcmin 15\farcs82$. We plot the optical position of this source corrected for proper motion  \citep{2013AJ....145...44Z} as a cross in Fig.~\ref{fig:emerlinmap}a, where the errors are indicated by the size of the cross, and find our radio detection in agreement within the errors. This signifies that the emission is stationary and coincides with the source and is not a faster moving knot of emission.

Due to the high resolution of e-MERLIN, we are unable to identify the $4\sigma_{\rm rms}$ source to the north in Fig.~\ref{fig:emerlinmap}a, although we suspect it may be extragalactic. In the absence of further data at similar resolution and sensitivity, we cannot say definitively and therefore do not discuss it further in this work.

Flux densities were extracted from the primary beam corrected maps using {\sc jmfit}. Errors on the flux densities are calculated as $\sigma = \sqrt{(0.1S_{\rm 5\,GHz})^2+\sigma_{\rm{rms}}^2+\sigma_{\rm{fit}}^2}$, where $0.1S_{\rm 5\,GHz}$ is 10\% absolute calibration error on the total 5\,GHz flux density $S_{\rm 5\,GHz}$, $\sigma_{\rm rms}$ is the rms noise and $\sigma_{\rm fit}$ is the fitting error returned from {\sc jmfit}. All errors are quoted to 1\,$\sigma$.

\section{Discussion}
\label{sec:disc}

\noindent \textbf{DG~Tau~A.} At 5\,GHz, we detect DG~Tau~A at $6\,\sigma_{\rm rms}$, and the source exhibits an asymmetric, compact and slightly elongated morphology consistent with the known direction of the outflow axis. We find a total flux density $S_{\rm t,5\,GHz}=312\pm78\,\mu$Jy and a peak flux density $S_{\rm p,5\,GHz}=160\pm38\,\mu$Jy. 

This source has been seen to possess a slightly rising radio spectrum with $\alpha=0.6$ \citep[Scaife et~al. 2012;][]{Lynch2013}, typical of collimated thermal jets or spherical stellar winds \citep{1986ApJ...304..713R}. However, when the total flux density from this work is compared with the nearly coeval JVLA A-configuration observations (Epoch 2011.46) at 8.5\,GHz from \citet{Lynch2013}, a spectral index $\alpha^{8.5}_{5}=2.4\pm0.5$ is found. Within the error this spectral index is possibly consistent with free-free emission, however $\alpha>2$ is generally considered unphysical for free-free emission. Although variability is expected to cause some difference in the measured flux densities between different observation epochs, we expect this is not the case here. In particular, the transit time across the source at maximum jet velocity is $\approx4$ months, which is longer than the amount of time between the JVLA and e-MERLIN epochs (1 month) and is therefore not the cause of the discrepancy. We suggest that the unphysical spectral index in this case arises predominantly from the mismatch in angular scales recovered by the two arrays. The longer baselines of the e-MERLIN array compared to those of the JVLA will reduce the sensitivity to larger scale structure. We expect source components $\geq\lambda/D_{\rm min}$ radians ($\sim1$\,arcsec for a minimum baseline of $D_{\rm min}=11.2$\,km) to be attenuated by more than 50\,percent of their flux, and therefore undetected. As a result, extrapolating between the e-MERLIN and JVLA data will give an inaccurate spectral index, which is indeed the case. 

To cross-check our commissioning data, we simulated the observations with the {\sc casa} simulator toolkit using a simple Gaussian flux distribution model of the JVLA data \citep{Lynch2013} as the sky model (deconvolved dimensions provided by C. Lynch, priv. comm.). Running {\sc simobserve}, we find that the angular scale and general morphology of DG~Tau~A from the simulated visibilities is in agreement with our observations. Additional observations with e-MERLIN at different frequencies are needed to constrain the spectral index on this scale, although with the improved sensitivity provided by the Lovell telescope and the full bandwidth of 2\,GHz, it may be possible to extract an instantaneous e-MERLIN spectral index from this source.

\noindent \textbf{DG~Tau~B.} We detect DG~Tau~B at $4\,\sigma_{\rm rms}$ within the e-MERLIN primary beam, with $S_{\rm t,5\,GHz}=150\pm63\,\mu$Jy and $S_{\rm p,5\,GHz}=113\pm38\,\mu$Jy. The morphology is jet-like and has a deconvolved PA of $125\degr$, consistent with the known jet axis \citep{1991A&A...252..740M}. A variable free-free component is suggested by the SED of DG~Tau~B, which could be explained by non-steady accretion/ejection \citep[e.g.][]{1995ApJS..101..117K}. There is contribution to the radio emission from the dust disc at higher radio frequencies (Scaife et~al. 2012), however the contribution at 5\,GHz for both DG~Tau~A and B should be negligible.

\subsection{Jet Opening Angle}
\label{sec:oa}

The strongest constraints on the jet collimation scale come from measurements of nearby T~Tauri stars with the Hubble Space Telescope \citep{1996ApJ...468L.103R} and ground-based adaptive optics, including DG~Tau~A \citep{2000A&A...357L..61D}. They show that jets from protostars appear resolved transversely and collimated as close as 35-50\,au from the central star, with initial opening angles of $\theta_{0}\approx20-30\degr$ which drop to only a few degrees beyond 50\,au. However, if the jet originates from a region within 1\,au of the star, it must have $\theta_{0}\geq45\degr$ to reach the observed width, and then undergo strong recollimation within 35-50\,au \citep{2002EAS.....3..147C}. 

We define the direction of the line of brightest emission to be the e-MERLIN jet direction of DG~Tau~A, which is consistent with that found in the optical. We take one dimensional cuts perpendicular to that direction with the {\sc AIPS} task {\sc slice} to determine the initial opening angle $\theta_{\rm eMERLIN}$ found with our e-MERLIN data. Each slice is fit with a Gaussian using {\sc slfit} and the FWHM of the Gaussian is then deconvolved from the synthesised beam to determine the FWHM of the jet at that distance. The peak of the emission in our 5\,GHz map (Fig.~\ref{fig:emerlinmap}a) is assumed to be the base of the jet, and the last slice of the jet was made at a distance of 0.07\,arcsec from this point and has a FWHM of 0.09\,arcsec (12.6\,au). As the emission at 5\,GHz terminates at this point, we identify it as the unity optical depth surface ($\tau=1$) and use the radius of the jet measured here to compute the mass-loss rate, see Section~\ref{sec:jetk}. 

\begin{figure}
\centerline{\includegraphics[width=0.5\textwidth]{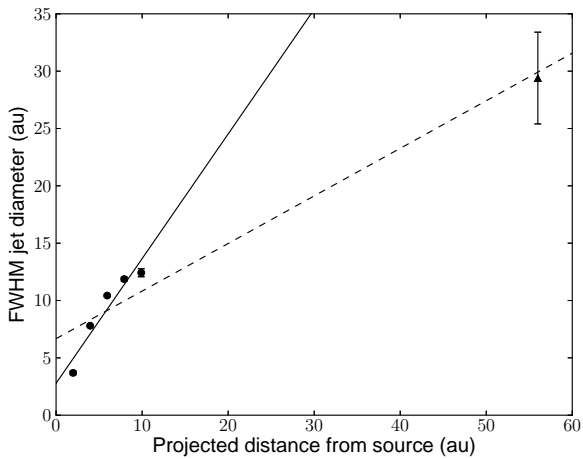}}
\caption{Deconvolved FWHM of the DG~Tau~A jet as a function of distance from the source. The measurements from these data are shown as filled circles and the errors are of order or smaller than the circles. The optical measurement at 56\,au from \citet{2004Ap&SS.292..643D} is shown as a triangle, and the error is an overestimate based on the range of optical measurements from these authors. The best-fitting regression to only the e-MERLIN measurements is shown as a solid line, and the dashed line represents the best fit when including the FWHM measurement at 56\,au. \label{fig:FWHMplot}}
\end{figure}

In Fig.~\ref{fig:FWHMplot}, we show the FWHM of the jet as a function of distance from the star, and include the optical measurement at 56\,au from \citet{2004Ap&SS.292..643D} in the plot along with our radio measurements. We fit two linear regressions: (1) to only the five e-MERLIN data points, and (2) to the e-MERLIN and 56\,au optical data points to calculate the average, constant opening angle for these ranges. From the linear fit to the e-MERLIN data we find an average, constant opening angle $\theta_{\rm eMERLIN}=71\degr$ (for distances $\simeq10$\,au), and when the optical point at 56\,au is included we find an angle of $30\degr$. For comparison, the average opening angle on scales of $\simeq100$\,au is $\approx14\degr$ based on data from \citet{2004Ap&SS.292..643D}.

If we derive the opening angles from only the first and last slices, we find $\theta_{\rm eMERLIN}=86\degr$ at 2\,au from the source and $\theta_{\rm eMERLIN}=64\degr$ at 10\,au. Many studies have shown that the initial opening angle is much larger than the average opening angle $\theta_{\rm a}$ further down the jet \citep[e.g.][]{1991A&A...252..740M, 1991A&A...252..733R}, indicating that jets are often only partially collimated on scales of tens of au and that a significant amount of collimation is achieved over the length of the jet. DG~Tau~A itself has also been shown to have a wider $\theta_{\rm a}$ compared with other CTTS, and it has been suggested that it is due to precession or contribution from strong bow-shock wings \citep{2000A&A...357L..61D}. \citet{1983ApJ...274L..83M} find an opening angle of $10\degr$ for a jet length of 8\,arcsec, showing further collimation on large scales. Our measurement of $\theta_{\rm eMERLIN}$ is much larger than that found further along the jet, suggesting that the DG~Tau~A jet starts initially as a poorly collimated wind close to the star and is recollimated further down the flow, in agreement with both disc wind \citep{2006A&A...453..785F} and X-wind \citep{2004IAUS..221..351S} theories.

\subsection{Mass-loss Rates}
\label{sec:jetk}

The mass-loss rate ($\dot{M}_{\rm j}$) is a vital parameter to derive as the mass ejection to accretion rate ratio is a constraining parameter for MHD launching models. The mass-loss rate for the blueshifted DG~Tau~A jet has been estimated by various authors using different techniques. Mass-loss rates ranging between $8\times10^{-9}-3\times10^{-7}$\,M$_{\sun}$\,yr$^{-1}$ have been estimated from observations at optical wavelengths \citep[][submitted]{1995ApJ...452..736H, 2000A&A...356L..41L, 2002ApJ...576..222B, 2008ApJ...689.1112C, 2011A&A...532A..59A, Maurri}. These estimates show discrepancies of almost two orders of magnitude depending on the adopted method for calculating $\dot{M}_{\rm j}$ due to high uncertainties on physical properties (e.g. jet radius) and extinction \citep{2007IAUS..243..203C, 2011A&A...532A..59A}. 

Using \textit{Herschel}/PACS observations, \citet{2012A&A...545A..44P} estimate $\dot{M}_{\rm j}=1.1\times10^{-7}$\,M$_{\sun}$\,yr$^{-1}$ from the [OI]\,63\,$\mu$m luminosity. If the ejected material is moving fast enough to produce a dissociated J-shock, then [OI] emission will be the dominant coolant in the postshock gas for temperatures of 100-5000\,K and is therefore a direct tracer of the mass flow into the shock. 

To directly compare $\dot{M}_{\rm j}$ found using our new e-MERLIN data with recent JVLA results, we follow the method described in \citet{Lynch2013}. Free-free emission depends on the observing frequency, the plasma temperature, and the linear emission measure. The optical depth at a given location in the jet can be estimated from source structure and the plasma temperature is known from optical line observations, therefore we can calculate the emission measure and further the electron density in the jet at that particular location. As the entire detected radio jet is the optically thick surface for the spatial resolution of e-MERLIN, we identify the observed termination of the radio jet with the unity optical depth surface ($\tau=1$) at 5\,GHz. We measure a distance of $\sim0.07$\,arcsec (9.8\,au projected for a distance of 140\,pc, see Section~\ref{sec:oa}) from the base of the jet to the $\tau=1$ surface. Assuming a mean temperature of $T=5000$\,K, the emission measure at 0.07\,arcsec is $EM=3.5\times10^{7}$\,pc\,cm$^{-6}$. This yields an electron density of $n_{\rm e}=8.6\times10^{5}$\,cm$^{-3}$ at this location in the jet.

We then use a simple jet density and cross-section model to compute $\dot{M}_{\rm j}$ in the ionised component. For a jet of density $\rho_{\rm j}$, cross-section $A_{\rm j}$, and velocity $V_{\rm j}$,  $\dot{M}_{\rm j} = \rho_{\rm j} A_{\rm j} V_{\rm j}$. Assuming a circular cross section of radius $r_{\rm j}$ and complete ionisation in the flow such that $\rho_{\rm j}=\mu m_{\rm p} n_{\rm e}$ (where $\mu=1.2$ is the mean atomic weight and $m_{\rm p}$ is the proton mass), the mass-loss rate can be re-written as
\begin{equation}
\left(\frac{\dot{M}_{\rm{j}}}{\rm{M}_{\sun}\,\rm{yr}^{-1}} \right) = 2.2\times10^{-11} \left(\frac{\mathit{n}_{\rm{e}}}{10^5\,\rm{cm}^{-3}}\right) \left(\frac{\mathit{r}_{\rm{j}}}{\rm{au}}\right)^{2} \left(\frac{\mathit{V}_{\rm j}}{100\,\rm{km\,s}^{-1}}\right).
\end{equation}
For an estimated jet radius of 6.3\,au at 0.07\,arcsec and an average radial velocity of 200\,km\,s$^{-1}$, we find $\dot{M}_{\rm j}=1.5\times10^{-8}$\,M$_{\sun}$\,yr$^{-1}$, within a factor of 3 of that found by \citet{Lynch2013}. This is an underestimate of the total mass-loss rate as it is only a measure of the ionised component. An ionisation fraction of $\sim14$\,per\,cent is found when comparing our value for $\dot{M}_{\rm j}$ with the total mass-loss rate from \citet{2012A&A...545A..44P}, which is consistent with previous measurements \citep[e.g.][]{2008ApJ...689.1112C}.

As jets are ultimately powered by accretion \citep{1990ApJ...354..687C, 1995ApJ...452..736H}, the ejection/accretion ratio is therefore a key parameter to constrain the jet acceleration mechanism and launch site \citep{2007IAUS..243..203C}. \citet{1995ApJ...452..736H} inferred a mean one-sided ratio $\dot{M}_{\rm j}/\dot{M}_{\rm acc}\backsimeq0.01$, however recent accretion rate estimates \citep{2011A&A...532A..59A} are an order of magnitude lower on average and therefore provide a ratio $10\times$ higher. 

\citet{2011A&A...532A..59A} find a possible range for the mass accretion rate $\dot{M}_{\rm acc}=(3\pm2)\times10^{-7}$\,M$_{\sun}$\,yr$^{-1}$ for the blueshifted jet of DG~Tau~A when comparing different estimates of this rate by different authors, and find $\dot{M}_{\rm j}/\dot{M}_{\rm acc}=0.04-0.4$ for the one-sided mass-loss to mass-accretion rate in the [FeII] emitting flow. \citet[submitted]{Maurri} find $\dot{M}_{\rm j}/\dot{M}_{\rm acc}=0.03-0.16$, and using the result for $\dot{M}_{\rm j}$ presented in this work, we find $\dot{M}_{\rm j}/\dot{M}_{\rm acc}=0.03-0.15$, compatible with the range of $0.01<\dot{M}_{\rm j}/\dot{M}_{\rm acc}<0.1$ predicted by MHD models for disc-winds \citep{2006A&A...453..785F} and $\dot{M}_{\rm j}/\dot{M}_{\rm acc}<0.3$ for X-winds \citep{2000prpl.conf..789S}. Multiplying by 2 to account for the redshifted jet yields $(2\dot{M}_{\rm j})/\dot{M}_{\rm acc}=0.06-0.3$ for the total ejection/accretion ratio, in agreement with other estimates \citep{2007IAUS..243..203C}.

\section{Conclusions}
\label{sec:conc}

In the case of DG~Tau~A, our results suggest that the jet starts initially as a poorly collimated wind and becomes collimated on scales of 50\,au, in line with MHD disc-wind theory. We find a large initial opening angle of $86\degr$ within 2\,au of the base of the jet which becomes smaller further out, a mass-loss rate of $\dot{M}_{\rm j}=1.5\times10^{-8}$\,M$_{\sun}$\,yr$^{-1}$ for the ionised component, and a total ejection/accretion ratio of $(2\dot{M}_{\rm j})/\dot{M}_{\rm acc}=0.06-0.3$. These results are in accord with predictions of MHD jet-launching models.

The improved sensitivity of e-MERLIN has provided the highest resolution images of the CTTS DG~Tau~A and the nearby Class~I protostar DG~Tau~B at cm-wavelengths. This work gives a preview on how e-MERLIN will contribute to the study of low-mass YSOs and their outflows, as observations of the inner 20\,au are needed to test different jet-launching scenarios. We have already probed these objects with a new level of angular resolution and sensitivity with e-MERLIN during the commissioning phase, and they will be further improved with the final array.

\section{ACKNOWLEDGEMENTS}
 
We thank John Bally for his constructive comments that helped to clarify these results. We thank the staff of the e-MERLIN/VLBI National Radio Astronomy Facility at Jodrell Bank Observatory for their assistance in the commissioning and operation of e-MERLIN. The reserach leading to these results received funding from the European Commission Severnth Framework Programme (FP/2007-2013) under grant agreement No 283393 (RadioNet3). REA would like to thank Anita Richards and Adam Avison of the University of Manchester for assistance with {\sc simobserve}, and Christene Lynch of the University of Iowa for thorough discussions on the topic of this paper. REA and TPR would also like to acknowledge support from Science Foundation Ireland under grant 11/RFP/AST3331.

\bibliographystyle{mn2e}
\bibliography{DGTAUbib}

\end{document}